\documentstyle[12pt]{article}

\renewcommand{\arraystretch}{1.3}
\relax

\begin{document}

\begin{titlepage}
 \null
\vskip 0.5in
\begin{center}
 \makebox[\textwidth][r]{UW/PT-92-01}
 \makebox[\textwidth][r]{DOE/ER/40614-16}
\makebox[\textwidth][r]{August 1992}
\vskip 0.35in
{\Large Jets at Hadron Colliders at Order
$\alpha_s^3$: A Look Inside}
 \par
 \vskip 0.6em
{\large
  \begin{tabular}[t]{c}
Stephen D.~Ellis \\
  \em Department of Physics \\
  \em University of Washington, Seattle, WA 98195, USA
   \and
Zoltan Kunszt \\
  \em Eidgenossosche Technische Hochshule \\
  \em CH-8093 Z\"urich, Switzerland
   \and
Davison E.~Soper \\
  \em Institute of Theoretical Science\\
  \em University of Oregon, Eugene, OR 97403, USA

\end{tabular}}
\par \vskip 1.0em
{PACS numbers 12.38, 13.87}
\vskip 1.0em

{\large\bf Abstract}
\end{center}
\quotation
Results from the study of hadronic jets in
hadron-hadron collisions
at order $\alpha_s^3$ in perturbation theory
are presented.  The focus is on various features of the
internal structure of jets.  The numerical
results of the calculation are compared with data
where possible and exhibit reasonable agreement.
\endquotation
\vspace{.2in}%
\centerline{PREPARED FOR THE U.S.~DEPARTMENT OF
ENERGY}
\medskip
\baselineskip 10truept plus 0.2truept minus 0.2truept
{\footnotesize
\noindent This report was prepared as an account of work
sponsored by the United States Government.  Neither the
United States nor the United States Department of Energy,
nor any of their employees, nor any of their contractors,
subcontractors, or their employees, makes any
warranty, express or implied, or assumes any legal liability
or responsibility for the product or process disclosed, or
represents that its use would not infringe privately-owned
rights.  By acceptance of this article, the publisher and/or
recipient acknowledges the U.S. Government's right to retain
a nonexclusive, royalty-free license in and to any copyright
covering this paper.}
\medskip

\end{titlepage}

\renewcommand{\baselinestretch}{1.5}

\pagestyle{plain}

\renewcommand{\arraystretch}{1.7}

Recent advances, both theoretical\cite{EKS,greco} and experimental\cite{CDF}%
, in the study of jet production in hadron collisions have made possible
detailed comparisons of theory with experiment. In the Standard Model our
general understanding of the high energy collisions of hadrons suggests that
jets arise when short distance, large momentum transfer interactions
generate partons (quarks and gluons) that are widely separated in momentum
space just after the hard collision. In a fashion that is not yet
quantitatively understood in detail these configurations are thought to
evolve into hadronic final states exhibiting collimated sprays of hadrons,
which are called jets. These jets are then the observable signals of the
short distance parton configurations.

This general qualitative picture is characteristic of both perturbative QCD
and the data. When one proceeds to a quantitative confrontation of theory
and experiment, a precise definition of a jet must be supplied, and measured
jet cross sections depend on the definition used. For instance, when one
defines a jet as consisting of all the particles whose momenta lie inside a
cone of radius $R$, then measured jet cross sections depend on $R$. On the
theoretical side, in an order $\alpha _s^3$ calculation a jet can consist of
two partons instead of just one. At this level, then, a jet can have
internal structure and jet cross sections calculated at order $\alpha _s^3$
will depend on the jet definition applied. Two questions arise. First, how
well does the dependence on the jet definition exhibited by the theoretical
jet cross section match that of the experimental jet definition? Second, how
well does the internal structure calculated at order $\alpha _s^3$ compare
to the internal structure of experimentally observed jets? We address these
questions in this Letter.

We consider the inclusive single jet cross section in $p\overline{p}$
collisions. This cross section is a function of the physical variables $s$,
the total energy, $E_T$, the transverse energy of the jet ($E_T=E\sin \theta
$), and $\eta $, the pseudorapidity of the jet ($\eta =\ln \cot (\theta /2)$%
) and, as suggested above, the definition of the jet. The theoretical
inclusive jet cross section also depends on the {\it unphysical}
renormalization/factorization scale $\mu $ and on the specific choice of
parton distribution functions, $f_{a/A}(x,\mu )$. In a Born level ($\alpha
_s^2$) calculation the dependence on the scale $\mu $ arises from both the
parton distribution functions and the parton scattering cross section
through the dependence of the latter on the strong coupling constant $\alpha
_s(\mu )$. At order $\alpha _s^3$ explicit factors of $\ln (\mu )$ appear
that serve to cancel a part of this $\mu $ dependence. If higher order
contributions were calculated, they would tend to eliminate, successively,
more of the $\mu $ dependence. At any fixed order in perturbation theory the
residual $\mu $ dependence acts as an estimator of the theoretical
uncertainty associated with the truncated perturbation series.

We find that the situation for the jet cross section at order $\alpha _s^3$
is a major improvement over the order $\alpha _s^2$ Born result both because
the $\mu $ dependence is reduced and because the cross section exhibits a
reasonable dependence on the jet definition. While the Born cross section
exhibits monotonic dependence on $\mu $, the higher order result is
relatively insensitive to the value of $\mu $ in a broad region near $\mu
\simeq E_T/2$. We estimate\cite{EKS} that the residual theoretical
uncertainty, as indicated by the residual $\mu $ dependence, is $\sim 10\%$.
The uncertainty in the cross section due to the current uncertainty in the
parton distribution functions is estimated\cite{EKS} to be somewhat larger,
$\sim 20\%$. In the perturbative calculations described here, the effects of
the long distance fragmentation processes and of the soft interactions of
the ``spectator'' partons are ignored. We estimate\cite{EKS} that these
uncalculated power suppressed effects constitute a correction of order $\sim
6\ {\rm GeV}/E_T$ to the cross section. Thus for jet $E_T$'s of order 100
GeV, as discussed here, the nonperturbative uncertainty is of the same order
or smaller than that due to perturbative effects.

At hadron colliders jets are typically defined\cite{accord} in terms of the
particles $n$ whose momenta $\overrightarrow{p_n}$ lie within a cone
centered on the jet axis ($\eta _J,\phi _J$) in pseudorapidity $\eta $ and
azimuthal angle $\phi $, $\left[ (\eta _n-\eta _J)^2+(\phi _n-\phi
_J)^2\right] ^{1/2}<R$. The jet angles ($\eta _J,\phi _J$) are the averages
of the particles' angles,%
\begin{eqnarray}
   \eta_J  & = & \sum_{n\in {\rm cone}} p_{T,n} \eta_n/E_{T,J} \,\, , \nonumber
 \\
   \phi_J  & = & \sum_{n\in {\rm cone}} p_{T,n} \phi_n/E_{T,J}
\end{eqnarray} with $E_{T,J}=\sum_{n\in {\rm cone}}p_{T,n}$. This process is
iterated so that the cone center matches the jet center ($\eta _J,\phi _J$)
computed in Eq. (1). It is important to note that this jet algorithm is not
yet fully defined since jets can overlap. In particular, it is possible for
the constraints above to be satisfied by configurations where some particles
are common to more than one cone. In the $\alpha _s^3$ calculation it is
possible for a one parton jet to lie within the cone of a two parton jet. In
the theoretical calculations described here\cite{EKS}, we use the rule that
only the two parton jets are included and the overlapping one parton jets
are discarded. The precise definitions used by the various experiments
differ to a greater or lesser extent from this form\cite{jetfind}.

While the Born cross section with only a single parton per jet is $R$ {\em %
independent}, the order $\alpha _s^3$ cross section can have 2 partons
inside a jet and is $R$ {\em dependent}. The theoretical expectation for the
$R$ dependence is shown in Fig. 1 along with results from CDF\cite{CDF}. The
inclusive single jet cross section is evaluated at $E_T=100$ GeV using the
HMRS(B)\cite{HMRS} parton distributions. Since dependence on $R$ is not
present in the Born cross section, this dependence is a lowest order result
at order $\alpha _s^3$, that is, $d\sigma /dR=0+{\cal O}(\alpha _s^3)$. One
therefore expects that, although the cross section itself is relatively $\mu
$ independent, its slope $d\sigma /dR$ will be quite strongly $\mu $
dependent. Just this behavior is indicated in Fig.~1 by the curves for the
cross section versus $R$ for 3 different $\mu $ values. (This figure is
essentially Fig. 3 of Ref. \cite{CDF} but with the correct theoretical
result. The fourth, dot-dash curve and the parameter $R_{sep}$ will be
explained below.  The values of the corresponding $R$-independent but
strongly $\mu$-dependent Born cross section are also indicated.)
A correlated feature is that the $\mu $ dependence of the
jet cross section changes as we vary $R$. While the order $\alpha _s^3$ jet
cross section is relatively independent of $\mu $ for $R\approx 0.7$, for
large $R$, $R > 0.9$, it is dominated by the order $\alpha _s^3$ real
emission process and becomes a monotonically decreasing function of $\mu $
much like the Born result. At small $R$, $R < 0.5$, the $\mu $ dependence
of the order $\alpha _s^3$ cross section is dominated by the {\em negative }%
contribution from the virtual correction associated with the collinear
singularity and becomes a monotonically increasing function of $\mu $. In
either regime we expect higher order corrections to be important so that the
usefulness of fixed order perturbation theory is compromised. Thus we
conclude that, at this order in perturbation theory, the results are most
stable for $R\approx 0.7$. It is precisely this size that was used in the
inclusive single jet cross section analysis published by CDF\cite{CDF}. In
some sense the perturbation theory is telling us that $R=0.7$ is the
``optimal size'' for a jet cone, at least from the standpoint of making
comparison with the order $\alpha _s^3$ result.

The comparison with the data in Fig. 1 suggests that, while the agreement
between theory and data for $R=0.7$ is quite good, the strong dependence on $%
R$ exhibited by the data favors a small $\mu $ value, {\it i.e., } larger $%
\alpha _s$ and more radiation. To make this comparison more quantitative we
can characterize both the data and the theory curves in terms of 3
parameters,
\begin{equation}
\sigma =A+B\ln R+CR^2\ .
\end{equation}
The parameterizations of the data and the theory for $\mu =E_T/2$ and $\mu
=E_T/4$ are indicated in the first three rows of Table 1 (the parameter $%
R_{sep}$ will be defined below). We see that the theoretical value for the $R
$-independent $A$ parameter is not too sensitive to $\mu $.  This is
expected since
it is a true one loop quantity, containing both the order $\alpha _s^2$
contribution and contributions from real and virtual graphs at order $\alpha
_s^3$. The $B$ and $C$ terms, however, are subject to larger theoretical
uncertainty since these terms express the $R$ dependence of the cross
section and this appears first at order $\alpha _s^3$. This is indicated by
the sensitivity to $\mu $ found in the Table. We can naively associate the $B
$ term with correlated (approximately collinear) final state parton emission
that is important near the jet direction and the $C$ term with essentially
uncorrelated initial state parton emission that is important far from the
jet direction.

The theoretical value of $A$ agrees quite well with the experimental value
of $A$. The agreement between the data and the $\mu =E_T/2$ theory for $C$
is also quite good, but the agreement for $B$ is {\em worse} than one would
expect. This suggests that for the $\mu =E_T/2$ theory, the amount of
initial state emission far from the jet direction is about right but that
there is not enough correlated radiation near the jet center. If we change $%
\mu $ to $E_T/4$, then the effective $\alpha _s$ is larger and there is more
radiation in all parts of phase space. Now $B$ is larger, although still
smaller than indicated by the data, while $C$ is larger than indicated by
the data.

To examine this issue further and to analyze the internal structure of jets
in detail, it is useful to consider the fractional $E_T$ profile, $F(r,R,E_T)
$ (we suppress the dependence on the jet direction ($\eta _J,\phi _J$)).
Given a sample of jets of transverse energy $E_T$ defined with a cone radius
$R$, $F(r,R,E_T)$ is the average fraction of the jets' transverse energy
that lies inside an inner cone of radius $r<R$ (concentric with the jet
defining cone). Said another way, the quantity $1-F(r,R,E_T)$ describes the
fraction of $E_T$ that lies in the annulus between $r$ and $R$. It is this
latter quantity that is most easily calculated in perturbation theory as it
avoids the collinear singularities at $r=0$. Computing the $E_T$ weighted
integral of the $p\overline{p}\rightarrow 3$ partons $+X$ cross section over
the annulus and normalizing to $E_{T,J}$ times the Born cross section yields
the order $\alpha _s$ contribution to $1-F$ (the numerator is purely order $%
\alpha _s^3$ while the denominator is purely order $\alpha _s^2$). The
result for $F$ is plotted in Fig. 2 versus the inner radius $r$ with $R=1.0$
for $E_T=100$ GeV and compared to preliminary CDF data\cite{CDFcone}. Again
curves for three choices of $\mu $ are exhibited. (The fourth, dot-dash
curve and the parameter $R_{sep}$ will be explained below.) As with the $R$
dependence of the cross section discussed above, $F$ is being calculated to
lowest nontrivial order and thus exhibits monotonic $\mu $ dependence. While
there is crude agreement between theory and experiment, the theory curves
are systematically below the data for all interesting values of $\mu $. This
situation suggests that the theoretical jets have too large a fraction of
their $E_T$ near the edge of the jet ($r\simeq R$).

We have seen that the
$R$ dependence and the $B$ parameter suggest
the importance of higher orders to increase the level of associated
radiation, at least near the center of the cone. At the same time our
detailed considerations of the parameter $C$ and of $F$ suggest that the
data favor a reduction of the $E_T$ fraction near the edge of the cone.
Although these conclusions seem contradictory, there may be a consistent
explanation based on a
detailed but important physical point concerning
how the jets are defined. We will
present a preliminary discussion of this point here and present a more
detailed study in a separate note.\cite{EKS2} The issue is that of {\em %
merging}, how close in angle should two partons be in order to be associated
as a single jet. In a real experiment such a situation is presumably
realized as two sprays of hadrons, each with finite angular extent due to
both fragmentation effects and real experimental angular resolution effects.
If the angular separation is large enough, there is a valley in the $E_T$
distribution between the two sprays and experimental jet finding algorithms
will tend to recognize this situation as two distinct jets. Recall that we
expect for jets of $E_T>100$ GeV that the angular extent of fragmentation
effects will be small compared to the defined jet cone sizes. However, the
theoretical jet algorithm we are using will merge two partons into a single
jet whenever it is mathematically possible. This includes the limiting
configuration when two equal transverse energy partons (each with $E_T/2$)
are just $2R$ apart. The calculation counts this as a single jet of
transverse energy $E_T$ with its cone centered
between the two partons, {\it %
i.e.,} centered on the valley. The treatment of this configuration in a real
experiment will depend in detail on the implementation of the jet algorithm.
To simulate the experimental
algorithm in a simple way we add an extra constraint
in our theoretical jet algorithm. When 2 partons, $a$ and $b,$ are separated
by more than $R_{sep}$($\leq 2R$), $R_{ab}=\left[ (\eta _a-\eta
_b^{})^2+(\phi _a-\phi _b)^2\right] ^{1/2}\geq R_{sep}$, we no longer merge
them into a single jet. The theoretical jet algorithm used above corresponds
to $R_{sep}=2R$ as noted in Table 1. As an example, the results of
calculating both the $R$ dependence and the $E_T$ fraction $F$ with $%
R_{sep}=1.3R$ and $\mu =E_T/4$ are illustrated by the dot-dash curves in
Figs. 1 and 2. Clearly the extra constraint of $R_{sep}$ has ensured that
there is approximately the observed fraction of $E_T$ near the edge of the
cone while the reduced $\mu $ value has increased the amount of associated
radiation near the center of the cone and produced a larger variation with $R
$. This conclusion is also verified by the last line in Table 1 where we
observe that, compared to the first line of theoretical results, $B$ has
increased while $C$ has remained the same. The values of both $B$ and $C$
that arise with these parameter choices are in reasonable agreement with the
data. The jet cross section itself is relatively insensitive to the
parameter $R_{sep}$, decreasing by $\leq $10\% as $R_{sep}$ is
reduced from $2R$ to $1.3R$ with fixed $\mu $ for $E_T=100$ GeV.

In summary, the
agreement between data and QCD perturbation theory at order $%
\alpha _s^3$ for the question of the dependence on the jet definition is a
vast improvement over the situation that obtained at the Born level. There
is good agreement between theory and experiment, at least for $E_T\geq 50$
GeV and $R$ near $0.7$. On the question of the detailed structure within
jets the qualitative agreement is good but there are important quantitative
issues that seem to be
dependent on the details of the implementation of the jet
definition, especially the question of jet merging. Further study, both
theoretical and experimental, is required to obtain a full understanding of
this problem\cite{EKS2}. This is particularly interesting since there is
some indication that such detailed internal jet structure can be invoked to
differentiate quark jets from gluon jets\cite{EKS2,Pump}.

\bigskip\


We thank the members of the CDF Collaboration's QCD group, and in particular
J. Huth and N. Wainer, for discussions concerning the CDF jet measurements.
This work was supported in part by U.S. Department of Energy grants
DE-FG06-91ER-40614 and DE-FG06-85ER-40224.

\relax
\def\pl#1#2#3{\ {\it Phys. Lett.} {\bf #1}, #2 (19#3)} \def\zp#1#2#3{\ {\it %
Zeit. Phys.} {\bf #1}, #2 (19#3)} \def\prl#1#2#3{\ {\it Phys. Rev. Lett.}
{\bf #1}, #2 (19#3)} \def\pr#1#2#3{\ {\it Phys. Rev.} {\bf D#1}, #2 (19#3)}
\def\np#1#2#3{\ {\it Nucl. Phys.} {\bf #1}, #2 (19#3)} \def\ib#1#2#3{\ {\it %
ibid.~}{\bf #1}, #2 (19#3)} \def\ar#1#2#3{\ {\it Ann. Rev. Nucl. Part. Sci. }%
{\bf #1}, #2 (19#3)}

\newpage

\begin{table}
\begin{center}

{\caption{3 parameter fits to data and calculated curves in Figs.~1,3 and
4}}
\begin{tabular}{||c|c|c|c||}\hline
 &A&B&C \\ \hline
CDF data\cite{CDFcone}&0.54&0.28&0.22\\
$\mu=E_T/2, R_{sep}=2R$&0.52&0.13&0.19\\
$\mu=E_T/4, R_{sep}=2R$&0.47&0.19&0.30\\
$\mu=E_T/4, R_{sep}=1.3R$&0.49&0.22&0.19\\ \hline
\end{tabular}
\end{center}
\end{table}

\bigskip\


\noindent {\bf {\Large Figure Captions}}

\begin{description}
\item[Fig. 1:]  Inclusive jet cross section versus $R$ for $\sqrt{s}=1800$
GeV, $E_T=100$ GeV and $0.1<|\eta |<0.7$ with $\mu =E_T/4,E_T/2,E_T$
compared to data from CDF\cite{CDF}; the dot-dash curve is explained in the
text. Also indicated for the three $\mu $ choices are the values of the
$R$-independent Born cross section.

\item[Fig. 2:]  $F(r,R,E_T)$ versus $r$ for $R=1.0$, $\sqrt{s}=1800$ GeV, $%
E_T=100$ GeV and $0.1<|\eta |<0.7$ with $\mu =E_T/4,E_T/2,E_T$ compared to
data from CDF\cite{CDFcone}; the dot-dash curve is explained in the text.
\end{description}

\end{document}